\documentclass[prl,aps,twocolumn,twoside,superscriptaddress]{revtex4}

\usepackage{amssymb}
\usepackage{amsmath}
\usepackage{amscd}
\usepackage{latexsym}
\usepackage{epsfig}
\usepackage{graphicx}
\usepackage{bbm}

\DeclareMathOperator{\Tr}{Tr}

\def\bra#1{{\langle #1 |  }}

\def\*{\star}
\def\({\left(}      
\def\){\right)}

\def\1{{\mathbf{1} }}

\def\ket#1{ | #1 \rangle}

\def\2pi{\hbox{$2\pi i$}}
\def\e#1{{\rm e}^{^{\textstyle #1}}}

\def\dsl{\raise.15ex\hbox{/}\kern-.57em\partial}
\def\Dsl{\,\raise.15ex\hbox{/}\mkern-.13.5mu D}

\newcommand{\proj}[1]{\ket{#1}\bra{#1}}

\def\2pi{\hbox{$2\pi i$}}
\def\e#1{{\rm e}^{^{\textstyle #1}}}

\def\dsl{\raise.15ex\hbox{/}\kern-.57em\partial}
\def\Dsl{\,\raise.15ex\hbox{/}\mkern-.13.5mu D}
\def\beq{\begin {equation}}
\def\eeq{\end {equation}}
\def\to{\rightarrow}

\def\12{{\textstyle \frac{1}{2}}}

\def\ss12{{\scriptstyle \frac{1}{2}}}
\def\32{{\scriptstyle \frac{3}{2}}}

\def\e2p#1{#1}

\begin{document}

\title{The exact cost of redistributing multipartite quantum states}
\author{Igor Devetak}
\affiliation{
Electrical Engineering Department,
University of Southern California,
Los Angeles, CA 90089, USA}
\author{Jon Yard}
\affiliation{Institute for Quantum Information, Caltech, Pasadena, CA 91125, USA}
\affiliation{Quantum Institute, CNLS, {CCS-3}, Los Alamos National Laboratory, Los Alamos, NM 87545, USA}

\begin{abstract} 
How correlated are two quantum systems from the perspective of a third?  We answer this by providing an optimal `quantum state redistribution' protocol for multipartite product sources.  Specifically, given an arbitrary quantum state of three systems, where Alice holds two and Bob holds one, we identify the cost, in terms of quantum communication and entanglement, for Alice to give one of her parts to Bob.  The communication cost gives the first known operational interpretation to quantum conditional mutual information.  The optimal procedure is self-dual under time reversal and is perfectly composable.  This generalizes known protocols such as the state merging and fully quantum Slepian-Wolf protocols, from which almost every known protocol in quantum Shannon theory can be derived.  
\end{abstract}

\maketitle
The statistical approach to information and the asyptotic analysis of the protocols which process it was pioneered by Claude Shannon \cite{shannon}.  He showed that the information content of a random variable $X$ with distribution $p(x)$ could be intuitively quantified by the \emph{Shannon entropy} 
$H(X) = -\sum_x p(x)\log_2 p(x)$.  More importantly, he \emph{operationally} justified this by proving that $H(X)$ is the minimum average number of bits required to faithfully represent independent instances of $X$ by any \emph{data compression protocol}, also proving that such protocols indeed exist.
 Shannon further defined the \emph{conditional entropy} as
$H(X|Y) = H(XY) - H(Y),$
which is also equal to the average entropy of $X$ given $Y$.
Conditional entropy  measures the information someone knowing only $Y$ would have to learn in order to know $X$ as well.
Its operational relevance was shown by Slepian and Wolf \cite{SW71} to be the minimum number of bits needed to describe $X$ to someone who knows $Y$. 
Shannon also introduced \emph{mutual information}
$I(X;Y) = H(Y) - H(Y|X)$
and  \emph{conditional mutual information} 
$I(X;Y|Z) = H(Y|Z) - H(Y|XZ)$,
each of which is interpreted as the information shared by $X$ and $Y$; the latter is measured from the perspective of someone knowing $Z$.  Mutual information plays a fundamental role in characterizing the capacity for a noisy channel to transmit information \cite{shannon}.  
Its conditional counterpart arises in the answers to many problems, such as in rate distortion with side information at the decoder \cite{WZ76}  and communication with side information at the encoder \cite{S58}.  It also appears in the analysis of degraded broadcast channels \cite{C72}.  All four of these quantities can easily be shown to be nonnegative.

In recent years, a quantum mechanical generalization \cite{BS98} of Shannon's theory has been developing
where a random variable is replaced with a quantum system $C$ with density matrix $\rho^C$.  The quantum analog of Shannon entropy is von Neumann entropy $H(C)_\rho = -\Tr\rho^C\log_2\rho^C$, which is the Shannon entropy of the eigenvalues of $\rho^C$.  While von Neumann's entropy preceded Shannon's by almost  twenty years, its operational interpretation was only found relatively recently by Schumacher \cite{Sch95}, who showed that a large number $n$ of quantum systems, identically prepared in the state $\rho^C$, could be compressed into a space of roughly $nH(C)$ \emph{qubits}, or two-level quantum systems. Here, a successful compression scheme is one which preserves the correlations $C$ shares with the rest of the world,  modeled by a reference system $R$.  The combined system is considered to be in any pure state $\ket{\psi}^{CR}$ satisfying $\rho^C = \Tr_R\proj{\psi}^{CR}$.  We then say that Alice holds a \emph{purification} of the reference $R$.   
The analogy can be continued, defining a quantum counterpart for each of Shannon's quantities by replacing Shannon with von Neumann entropies.  \emph{Quantum mutual information} \cite{CA97a,BSST99} $I(A;B)$ can be considered as a measure of correlations between $A$ and $B$.  It plays a remarkably similar role as its classical counterpart, describing the classical capacity of a noisy quantum channel in the presence of free entanglement \cite{BSST99,BSST02} (see also  \cite{GPW05} for a thermodynamical interpretation). 
On the other hand,  \emph{quantum conditional entropy} $H(A|B)$ \cite{CA97a} and \emph{quantum conditional mutual information} (QCMI) $I(A;B|C)$ are less like their classical counterparts, as they cannot generally be viewed as averages.  Furthermore, $H(A|B)$ can be negative; $-H(A|B)$ is often referred to as the \emph{coherent information} \cite{SN96}, which plays a role in characterizing the capacity of a quantum channel for transmitting quantum information \cite{L96,S02,D05}.   The operational task of \emph{state merging} \cite{HOW05} gives meaning to $H(A|B)$ where, depending on its sign, it corresponds to the rate at which entanglement is either consumed or generated while transferring $A$ to someone already holding $B$. 
On the other hand, QCMI can be shown to be nonnegative. Unlike the classical case, this amounts to a theorem, known as \emph{strong subadditivity} of quantum entropy,  whose original proof \cite{LR73} relies on nontrivial tools from matrix analysis.  More recently, operational proofs have been found \cite{HOW05,GPW05}, and we will see that our protocol leads to yet another such proof.   Strong subadditivity is correspondingly powerful; it underlies virtually every known bound in quantum information theory.  
Despite its central role, a direct operational interpretation of QCMI on an arbitrary state has been conspicuously absent, although it has arisen in operational interpretations for certain restricted classes of underlying states \cite{CW04}.  Such a general interpretation is provided in this paper.

\begin{figure}
\center
\includegraphics[scale=.5]{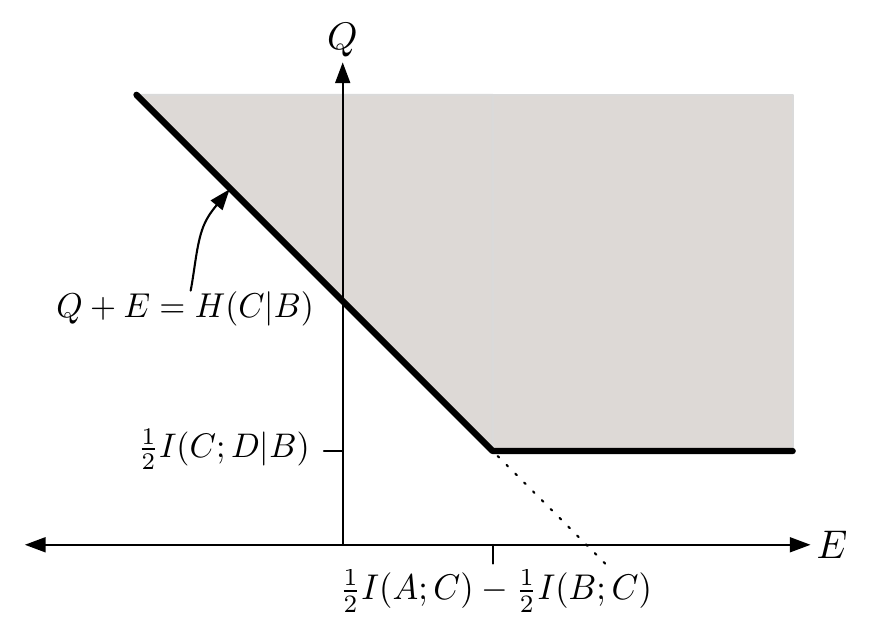}
\caption{Region of achievable cost pairs, together with the time-reversed optimal cost pair, assuming $I(C;A) > I(C;B)$.}
\label{figure:region}
\end{figure}

The ability to send qubits from Alice to Bob is a resource, and Schumacher's theorem tells how much of it is needed to transfer $C$.   A weaker resource is entanglement, because it can be established by by sending qubits.  A ``standard unit" of entanglement is called an \emph{ebit} and consists of a single EPR pair 
$\ket{\Psi_{\!+}} = \frac{1}{\sqrt{2}}\big(\ket{00} + \ket{11}\big)$
shared between Alice and Bob.   By quantum teleportation \cite{BBCJPW93}, an ebit can be used to send a qubit, provided that two classical bits are sent as well.  In the absence of classical communication, ebits are not helpful for moving $C$ from Alice to Bob.  They are helpful, however, in a variant of Schumacher's scenario in which  Alice and Bob have some side information.   We model this with four systems in the state $\ket{\psi}^{ACBR}$ .  We begin by assuming that Alice holds $AC$, Bob holds $B$, while the reference $R$ is unavailable to both parties.  Alice's and Bob's task is to \emph{redistribute} the quantum information so that it instead Bob who holds $C$ as follows:
\[(AC)^{\text{Alice}} (B)^{\text{Bob}} \to (A)^{\text{Alice}}(CB)^{\text{Bob}}.\]
To achieve this, Alice and Bob may perform local operations, Alice may send Bob qubits, and Alice and Bob may consume or generate entanglement.  
In particular, no classical communication is allowed, beyond what can be encoded in qubits.  We allow the entanglement cost to be any real number, interpreting positive and negative values as in state merging.  Our main result is that there exists a protocol --  \emph{quantum state redistribution} -- allowing Alice to transfer $C$ to Bob at a cost of $Q$ qubits and $E$ ebits if and only if 
\begin{eqnarray*}
Q &\geq& \12I(C;R|B) \\
Q + E &\geq& H(C|B).
\end{eqnarray*} 
This gives the first direct operational interpretation of QCMI on an \emph{arbitrary} state.
 Simultaneously minimizing $Q$ and $Q+E$ leads to the \emph{optimal cost pair}
\begin{eqnarray}
Q&=& \12 I(C;R|B) \label{redistpairQ}\\
E &=& \12 I(C;A) - \12 I(C;B). \label{redistpairE}
\end{eqnarray}
This pair corresponds to the corner point of the region in FIG.~\ref{figure:region}.
As with Schumacher compression, this result is to be understood in the limit of many identical copies.  Let us now point out some remarkable features of this result.

\begin{figure}
\center
\includegraphics[scale=.4]{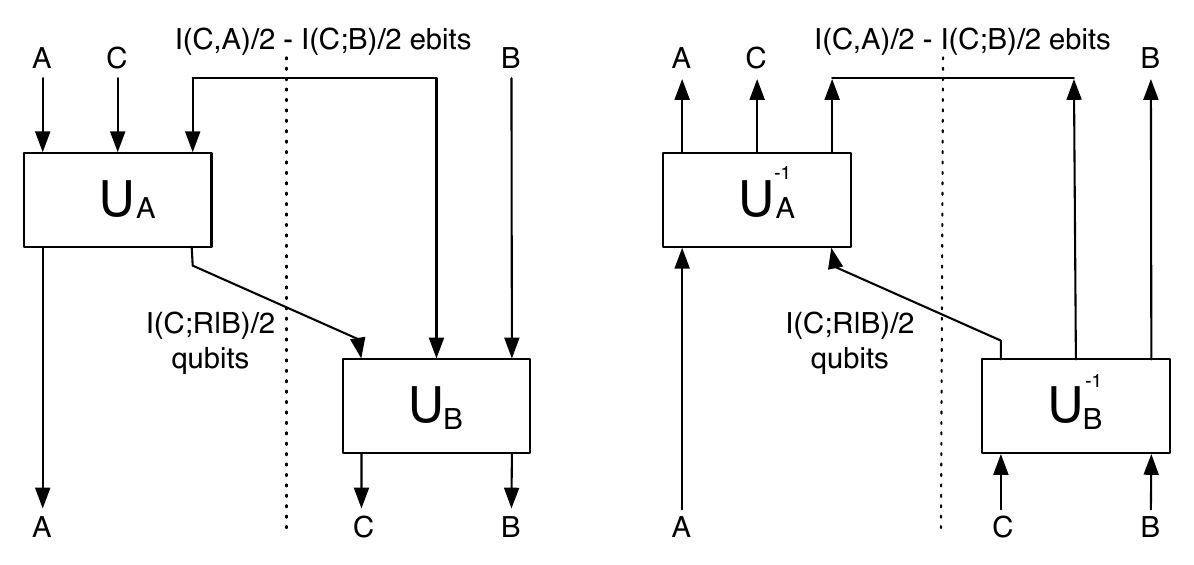}
\caption{Left: Unitary quantum state redistribution protocol in which Alice redistributes $C$ to Bob while \emph{consuming} entanglement (assuming that $I(C;A) \geq I(C;B)$).  Right: Corresponding time-reversed process where Bob redistributes $C$ to Alice, this time \emph{generating} the same amount of entanglement.}
\label{figure:circuit}
\end{figure}

\paragraph*{Self-duality under time reversal:} 
As illustrated in FIG.~\ref{figure:circuit}, our protocol can be implemented \emph{unitarily} -- if entanglement is consumed by the protocol, reversing those unitaries leads to a protocol which instead sends $C$ from Bob to Alice, while \emph{generating} the same amount of entanglement.  
Perhaps surprisingly, this symmetry is also evident in the \emph{optimal} cost pairs: switching $A$ and $B$ reflects the optimal cost pair about the $Q$-axis (see Figure~\ref{figure:region}).  Thus, switching $A$ and $B$ changes the sign of $E$ in (\ref{redistpairE}), but has no effect on the expression (\ref{redistpairQ}) for $Q$ because the identity $I(C;R|B) = I(C;R|A)$ holds on every pure state $\ket{\psi}^{ABCR}$.   In fact, our protocol can be considered as providing an explaination for why this identity should be true.

\paragraph*{Perfect composability:} Suppose that Alice wants to transfer a composite 
system $CD$ to Bob.  An optimal strategy is for Alice to treat
$CD$ as a single system, sending them both simultaneously using our protocol. 
The optimal cost pair for this is 
\begin{eqnarray*}
Q&=&\12I(CD;R|B) \\ 
E &=& \12 I(CD;A) - \12I(CD;B).
\end{eqnarray*}
What if she sends the systems successively?
The optimal cost for first transferring $D$ is
\begin{eqnarray*}
Q_D &=& \12I(D;R|B) \\
E_D &=&  \12I(D;AC) - \12I(D;B).
\end{eqnarray*}
Since Bob now has $D$, the remaining cost for sending $C$ is
\begin{eqnarray*}
Q_C &=&  \12 I(C;R|DB) \\
E_C  &=& \12 I(C;A) -\12 I(C;DB).
\end{eqnarray*}
Simple algebraic manipulations then show that $Q = Q_C + Q_D$ and $E = E_C + E_D$! This feature
parallels successive refinement in classical rate-distortion theory \cite{CE91},
only here the Markov condition is absent.

\paragraph*{Applications:}
Consider the following illustrative examples and applications of state redistribution:

\emph{(1) Four-party cat state:}  The optimal cost pair for the state 
$\frac{1}{\sqrt{2}}\big(\ket{0000} + \ket{1111}\big)^{ACBR}$ is  $Q = E = 0$.  To redistribute $C$ from Alice to Bob, Alice applies the local isometry $\ket{0}^A\bra{00}^{AC}+ \ket{1}^A\bra{11}^{AC}$, after which Bob applies  $\ket{00}^{CB}\bra{0}^{B}+ \ket{11}^{CB}\bra{1}^{B}$.

\emph{(2) Four-party W state:} If the global state is $\frac{1}{2}\big(\ket{1000} + \ket{0100} + \ket{0010} + \ket{0001}\big)^{ACBR},$
we obtain $Q\approx .38$ and $E=0$ for the optimal cost pair.   For comparison, a compress-and-send strategy which ignores the side information requires roughly $.81$ qubits.  

\emph{(3) States saturating strong subadditivity:}
The states which require a zero rate of communication to redistribute $C$ are precisely those which saturate strong subadditivity ($I(C;R|B) =0$), and are thus locally equivalent to a state of the 
form \cite{HJPW04}
\[\sum_x \sqrt{p_x}\ket{x}^{A'}\ket{x}^{B'}\ket{\phi_x}^{A_CB_CC}\ket{\varphi_x}^{A_RB_RR}.\]
The entanglement cost for such states is 
$\sum_x p_x  \big(H(B_C)_{\phi_x} - H(A_C)_{\phi_x}\big).$
Another optimal strategy is thus to coherently concentrate \cite{BBPS96} the $A_CC\big|B_C$ entanglement while diluting \cite{LP99} the $A_C\big|CB_C$ entanglement in the individual states $\phi_x^{A_CB_CC}$.  

\emph{(4) State merging:}
Our state redistribution protocol allows for a deeper understanding of state merging \cite{HOW05,HOW05b}.  By adding the additional resource of free classical communication, we recover their result that the cost, in ebits, for merging $C$ to $B$ is equal to $H(C|B)$.  Accounting for transmitted bits as well, state merging considers $A$ to be part of the reference, requiring that $I(RA;C)$ bits be sent per copy of $C$ merged.
By our result, the classical communication cost is reduced to $I(R;C|A)\leq I(RA;C)$, which can be shown to be optimal by an argument similar to the one we give for our protocol.  Thus, QCMI can also be regarded as the classical communication cost for state transfer in the presence of unlimited entanglement.

\emph{(5) Fully quantum Slepian-Wolf (FQSW):}
A special case of our result is when Alice has no side information.  An optimal strategy for this scenario has been found previously and called the \emph{fully quantum Slepian-Wolf} protocol \cite{D05b,ADHW06}, which can transfer $C$ from Alice to Bob using $Q$ qubits and $E$ ebits if and only if 
\begin{eqnarray*}
Q &\geq& \12 I(C;R) \\
Q + E &\geq& H(C|B).
\end{eqnarray*}

\emph{(6) Fully quantum reverse Shannon (FQRS):}
If it is instead Bob who lacks side information, we obtain the previously studied \cite{D05b,ADHW06} fully quantum reverse Shannon protocol as a special case of our result. 
Here, the required costs for transferring $C$ to Bob are given by 
\begin{eqnarray*}
Q &\geq& \12 I(C;R) \\
Q+E &\geq& H(C).
\end{eqnarray*}  
The optimal cost pair is dual to that of FQSW under time reversal \cite{D05b}.

\paragraph*{The protocol:}
Here we describe the proof that our protocol exists;  for a more detailed treatement see \cite{DY06a}.
First note that by the FQRS protocol, Alice can use $\12 I(C;RB)$ qubits and $\12 I(C;A)$ ebits to simulate the isometry which moves $C$ to Bob, while keeping $A$ to herself.  In order to take advantage of Bob's side information, Alice can use a modification of that protocol which also transmits $I(C;B)$ bits per copy of $C$ which is moved \cite{HHHLT01,BSST02}.  It is furthermore possible to make the classical communication \emph{coherent} \cite{H04,DHW04} in the following sense.  We say that Alice sends a \emph{coherent bit} \cite{H04} to Bob if she applies an isometry $\ket{x}^A \mapsto \ket{x}^A\ket{x}^B$ to a qubit in her possession, where Bob has $B$ and $x=0,1$.   Asymptotically, two coherent bits can be used to send a qubit and to generate an ebit \cite{H04}, so Alice can send an \emph{additional} $\frac{1}{2}I(C;B)$ qubits, while generating the same number of ebits with Bob.  This leads to a catalytic scenario, where extra ebits and qubits are needed to start the protocol, but are returned after completion.  The dependence on the catalysts can be eliminated using methods in \cite{DHW05}.   Subtracting the resources generated by the protocol from those which were invested yields the optimal cost pair.  On the other hand, the optimality of our protocol is shown in \cite{LD06} to follow from that of FQSW by subtracting the optimal costs for Alice to send only $A$ from the costs for Alice to send $AC$.

\paragraph*{Discussion:}
For an arbitrary pure state $\ket{\psi}^{ABCR}$, we have determined the communication and entanglement resources which are necessary and sufficient for Alice and Bob, who respectively hold $A$ and $B$, to transfer $C$ between themselves while retaining the purity of the global state.  The optimal communication cost gives the first operational interpretation of QCMI on an \emph{arbitrary} state and also gives a natural interpretation to the pure state identity $I(C;R|A)=I(C;R|B)$: the correlations between $C$ and $R$ look the same from each of Alice's and Bob's perspectives.  Because of this, the communication cost is symmetric under time-reversal.  On the other hand, the optimal entanglement cost was shown to be \emph{antisymmetric} under time-reversal, so that if ebits are consumed to move $C$ one way, the same number are generated while moving it back.

There is a formal time-reversal duality between FQSW and FQRS \cite{D05b}.  We showed that our protocol is \emph{self-dual} in the same sense, while incorporating both results as special cases.  Interestingly, our coding theorem is based on a generalization of that from FQRS, while both the coding theorem and the converse from FQSW are used for our converse.

A corollary of our main result is a direct operational proof of strong subadditivity.  Ours differs from other such operational proofs  \cite{HOW05,GPW05} because it 
does not even rely on the \emph{subadditivity} of entropy, i.e.\ $H(A)\geq H(A|B)$.  
Before removing the dependence on catalyst channels, our protocol cannot simulate more qubit identity channels
than were initially provided, so that positivity of the overall qubit cost (QCMI) is evident.

Because our protocol involves an arbitrary four-partite pure state, 
it can be applied as a fundamental primitive for all multi-party state redistribution problems.  Indeed, whenever there is a sender (Alice) and a receiver (Bob), there
are four natural subsystems: the system $A$ which stays with Alice, the system $B$ which Bob
already has, the system $C$ which is being communicated, and the rest of the world $R$. Even if there are many more parties, each particular round of communication fits into our setting.  For instance, suppose Alice holds $AC_A$, Bob has $BC_B$ and Charlie holds $C$, while all systems are purified into a reference system $R$.  If the goal is to transfer $C_A$ and $C_B$ to Charlie, direct application of our result gives a four-dimensional region of achievable costs $(Q^{A\to C},Q^{B\to C},E^{AC},E^{BC})$, generated by two corner points, each corresponding to a different order in which Charlie receives the systems $C_A$ and $C_B$.  It is likely that other strategies, such as where $C_A$ and $C_B$ are split into multiple subsystems and are sent to Charlie in various orders, would lead to even larger achievable regions. Furthermore, it is known \cite{ADHW06} that FQSW, when combined with teleportation \cite{BBCJPW93} and superdense coding \cite{BW92}, recovers virtually every known quantum Shannon-theoretic protocol.  We expect even more from state redistribution and are currently investigating its further implications for constructing more complex protocols and for understanding the structure of multipartite quantum states. 

\paragraph{Acknowledgements: }
The authors are grateful to Toby Berger for urging them to find
a quantum analogue of successive refinement, which partly inspired this result.
They also thank Andrew Childs and Aram Harrow for comments and acknowledge funding from grants through the NSF.  JY's research at LANL is supported by the Center for Nonlinear Studies (CNLS), the Quantum Institute and the LDRD program of the U.S. DOE.

\bibliography{qcmi}

\end{document}